\documentclass[journal]{IEEEtran}
\usepackage[utf8]{inputenc}
\usepackage{amsmath}
\usepackage{amsfonts}
\usepackage{amssymb}
\usepackage{mathtools}
\usepackage{amsthm}
\usepackage{graphicx}
\usepackage[caption=false, font=footnotesize]{subfig}
\usepackage{tikz}
\usetikzlibrary{arrows,shapes,positioning}

\newtheorem{theorem}{Theorem}

\newtheorem{lemma}{Lemma}

\newtheorem{corollary}{Corollary}
\newtheorem{remark}{Remark}

\newcommand{\E}{\operatorname{E}}

\newcommand{\Cov}{\operatorname{Cov}}

\title{Communication Through a Large Reflecting Surface With Phase Errors}
\author{Mihai-Alin~Badiu and Justin~P.~Coon,~\IEEEmembership{Senior Member,~IEEE}
\thanks{The authors are with the Department of Engineering Science, University of Oxford, Parks Road, Oxford, OX1 3PJ, UK, (email: mihai.badiu@eng.ox.ac.uk; justin.coon@eng.ox.ac.uk).}
\thanks{This material is based upon work supported by, or in part by, the U. S. Army Research Laboratory and the U. S. Army Research Office under contract/grant number W911NF-19-1-0048.}
}

\begin{document}

\maketitle

\begin{abstract}
Assume the communication between a source and a destination is supported by a large reflecting surface (LRS), which consists of an array of reflector elements with adjustable reflection phases. By knowing the phase shifts induced by the composite propagation channels through the LRS, the phases of the reflectors can be configured such that the signals combine coherently at the destination, which improves the communication performance. However, perfect phase estimation or high-precision configuration of the reflection phases is unfeasible. In this paper, we study the transmission through an LRS with phase errors that have a generic distribution. We show that the LRS-based composite channel is equivalent to a point-to-point Nakagami fading channel. This equivalent representation allows for theoretical analysis of the performance and can help the system designer study the interplay between performance, the distribution of phase errors, and the number of reflectors. Numerical evaluation of the error probability for a limited number of reflectors confirms the theoretical prediction and shows that the performance is remarkably robust against the phase errors.
\end{abstract}


\section{Introduction}
The wireless propagation environment is traditionally viewed as a detrimental medium of communication that distorts the transmitted signals in an uncontrollable manner. Wireless systems have improved consistently over the years through the development of increasingly sophisticated transmission and reception techniques that counteract the undesirable effects of propagation. An alternative to this conventional reactive approach is currently gaining interest, as several works propose to achieve a level of control over the propagation of electromagnetic waves by placing in the environment `smart' passive devices that manipulate the impinging waves according to a desired functionality, such as reflection in an intended direction, full absorption, polarization, or filtering~\cite{Tan2016,Liaskos2018,Renzo2019}. Furthermore, by endowing such devices with interconnection capability, they can be introduced in control loops that enable adaptation to the changing conditions of the wireless network, thus creating a smart radio environment. Examples of reconfigurable devices that can manipulate EM waves include tunable reflect-arrays~\cite{Hum2007,Tan2016}, frequency-selective surfaces~\cite{Subrt2012},  or meta-surfaces consisting of a thin layer of metamaterials~\cite{Cui2014}. The latter have particularly attractive properties and are expected to be a key enabler of smart wireless environments~\cite{Liaskos2018,Renzo2019}. 

The focus of this paper is on the concept of a large reflective surface (LRS) which, irrespective of the building technology, can be abstracted into an array of passive reflector elements that induce adjustable phase shifts on the reflected signals. An LRS placed somewhere between a source and a destination can greatly enhance the communication between the two when the phase shifts are appropriately configured such that the reflected waves combine coherently at the destination~\cite{Tan2016,Huang2018,Wu2018,Basar2019}. Most of the studies assume the phase shifts induced by the composite propagation channel through the LRS are perfectly known and the reflector phases are set to the ideal values, with no errors. However, perfect phase estimation or high-precision configuration of the reflection phases is unfeasible. In~\cite{Huang2018b,Wu2019}, the LRS-based system is optimized over low-resolution phase shifts.

In this paper, we study the communication through an LRS whose induced phase shifts deviate from the ideal values according to a generic probability distribution.  This set-up could, for example, represent imperfect phase estimation, quantized reflection phases, or both. We show that the transmission through a large number $n$ of imperfect reflectors is equivalent to a point-to-point communication over a Nakagami fading channel with the following characteristics: the diversity order grows with $n$, the average SNR scales with $n^2$ relative to the average single-reflector SNR, and both parameters are attenuated by the phase uncertainty. Numerical results for a limited number of reflectors validate the theoretical prediction and show that accurate knowledge or representation of the ideal phase shifts is not necessary to closely approach the ideal performance. 

\section{System Model}
The considered system comprises a single-antenna source $\mathrm{S}$, a single-antenna destination $\mathrm{D}$ and an LRS with $n$ reflector elements $\mathrm{R}_1,\ldots,\mathrm{R}_n$, which assist the communication between $\mathrm{S}$ and $\mathrm{D}$. Since we want to study the effect of the phase errors at the LRS, we restrict our focus on the transmission through the LRS only and assume there is no direct link between $\mathrm{S}$ and $\mathrm{D}$. Assuming slow and flat fading, let $H_{i1}$ and $H_{i2}$ be the complex fading coefficients of the $\mathrm{S}$-to-$\mathrm{R}_i$ and, respectively, $\mathrm{R}_i$-to-$\mathrm{D}$ channels. The reflectors are sufficiently spaced apart such that all the $2n$ coefficients are mutually independent. Furthermore, they have unit power and average magnitudes
\begin{equation}\label{eq:absH}
    a_1 = \E[|H_{i1}|]<1\quad\text{and}\quad a_2=\E[|H_{i2}|]<1,
\end{equation}
for all $i=1,\ldots,n$. In the following, we use
\begin{equation}
    a=\sqrt{a_1 a_2}.
\end{equation}

Denote by $\phi_i$ the phase shift induced by $\mathrm{R}_i$. The signal received by the destination over the equivalent baseband channel is\footnote{The signal is normalized by the standard deviation of the receiver noise.}  
\begin{equation}\label{eq:RX_sig}
    Y = \sqrt{\gamma_0}\sum_{i=1}^n H_{i2} \, e^{j\phi_i} H_{i1} X + W,
\end{equation}
where $X$ is the transmitted symbol with mean zero and $\E[X^2]=1$, $W\sim \mathcal{CN}(0,1)$ models the normalized receiver noise and $\gamma_0$ is the average SNR when only one reflector is present and accounts for path loss, shadowing and reflection loss. 

Ideally, given the phases $\arg(H_{i1})$ and $\arg(H_{i2})$, the phase $\phi_i$ is set so as to cancel the overall phase shift $\arg(H_{i1})+\arg(H_{i2})$, which maximizes the SNR at the receiver~\cite{Basar2019}. Assuming  the phase shifts induced by the channels are not estimated perfectly and/or the desired phases cannot be set precisely (e.g., when only a discrete set of phases is technologically possible), we model the deviation of $\phi_i$ from the ideal setting by the phase noise $\Theta_i$, which is randomly distributed on $[-\pi,\pi)$ according to a certain  circular distribution~\cite{MardiaJupp2000}. 

Now, we express the received signal as
\begin{equation}\label{eq:RX_sig2}
    Y = n\sqrt{\gamma_0} H X + W,
\end{equation}
where
\begin{equation}\label{eq:H}
    H = \frac{1}{n}\sum_{i=1}^n |H_{i1}||H_{i2}| e^{j\Theta_i} \in\mathbb{C}.
\end{equation}
All the variables in the r.h.s. of~\eqref{eq:RX_sig2} and~\eqref{eq:H} are mutually independent. We assume $\Theta_i$, $i=1,\ldots,n$, are i.i.d. with common characteristic function expressed as a sequence of complex numbers $\{\varphi_p\}_{p\in\mathbb{Z}}$, 
\begin{equation}\label{eq:trig_mom}
    \varphi_p=\E[e^{j p\Theta}],
\end{equation}
which are also called trigonometric (or circular) moments~\cite{MardiaJupp2000}. Note that $|\varphi_p|\leq 1$, for all $p\in\mathbb{Z}$; for $|p|\geq 1$ equality holds if, and only if, the distribution is degenerate (Dirac delta). A distribution that is broad on the circle (signifying high uncertainty) gives small magnitudes of the trigonometric moments (for the uniform distribution, $\varphi_p=0$ for $p\geq 1$). It is reasonable to assume that the distribution of the error $\Theta_i$ has mean direction zero ($\arg\E[e^{j\Theta_i}]=0$) and its pdf is symmetric around zero. Due to the symmetry, the characteristic function takes real values, i.e., $\varphi_p\in\mathbb{R}$, for all $p\in\mathbb{Z}$. 

The receiver performs ideal coherent detection based on the sufficient statistic
\begin{equation}
    e^{-j\arg(H)}Y = n\sqrt{\gamma_0} |H| X + W',
\end{equation}
where $W'=e^{-j\arg(H)}W\sim\mathcal{CN}(0,1)$.

\section{Equivalent Scalar Fading Channel}
In this section, we find an equivalent representation of the transmission through an LRS as a point-to-point fading channel. First, we provide the following result.
\begin{lemma}\label{lm:H}
For large $n$, the coefficient $H$ given by~\eqref{eq:H} has a (non circularly symmetric) complex normal distribution. Its real and imaginary parts, $U=\Re(H)$ and $V=\Im(H)$, are independent, and  $U\sim\mathcal{N}(\mu,\sigma_U^2)$ and $V\sim\mathcal{N}(0,\sigma_V^2)$ with
\begin{align}
    \mu &= \varphi_1 a^2\\
    \sigma_U^2 &= \frac{1}{2n}(1+\varphi_2-2\varphi_1^2 a^4)\\
    \sigma_V^2 &= \frac{1}{2n}(1-\varphi_2).
\end{align}
\end{lemma}
\begin{IEEEproof}
See Appendix~\ref{app:1}.
\end{IEEEproof}
 
The ideal case where the phase shifts induced by the LRS-augmented channel are perfectly estimated and the reflector phases can be set precisely is instantiated as follows. 
\begin{corollary}\label{cor:H_ideal}
When there are no phase shift errors,  i.e., $\Theta_i\equiv 0$, for all $i=1,\ldots,n$, the coefficient $H$ in~\eqref{eq:H} is real and distributed as $\mathcal{N}\left(a^2,(1-a^4)/n\right)$.
\end{corollary}
\begin{IEEEproof}
Since the phase errors are zero, we have $\varphi_1 = \varphi_2 = 1$. This implies that $\Im(H)\equiv 0$ in Lemma~\ref{lm:H} and the result follows. 
\end{IEEEproof}
The ideal case of no phase errors and Rayleigh fading (i.e., $a=\sqrt{\pi}/2$) is studied in~\cite{Basar2019}, which Corollary~\ref{cor:H_ideal} recovers. 

\begin{corollary}\label{cor:H_no_knowledge}
When the phase shift errors are uniformly distributed on $[-\pi,\pi)$, which corresponds to a complete lack of knowledge about the phases of $H_{i1}$ and $H_{i2}$,  $i=1,\ldots,n$, the coefficient $H$ in~\eqref{eq:H} has a circularly-symmetric complex normal distribution with mean zero and variance $1/n$, which means the equivalent channel resembles Rayleigh fading.
\end{corollary}
\begin{IEEEproof}
For the uniform circular distribution, $\varphi_1 = \varphi_2 = 0$ and the result follows from Lemma~\ref{lm:H}. 
\end{IEEEproof}

When $\varphi_1>0$, even though $U$ and $V$ are Gaussians with non-zero and, respectively, zero means, the magnitude $|H|=\sqrt{U^2+V^2}$ does not have a Rice distribution because $\sigma_U^2\neq\sigma_V^2$ in general. We obtain the following result instead. 
\begin{theorem}\label{th:absH}
For large $n$ and $\varphi_1>0$, the magnitude of $H$ in~\eqref{eq:H} has a Nakagami distribution with pdf
\begin{equation}\label{eq:pdf_absH}
    f_{|H|}(x) = \frac{2m^m}{\Gamma(m)\mu^{2m}} x^{2m-1}\exp\left(-\frac{m}{\mu^2}x^2\right),
\end{equation}
where the fading parameter $m$ is given by
\begin{equation}
    m = \frac{n}{2}\,\frac{\varphi_1^2 a^4}{1 + \varphi_2 - 2\varphi_1^2 a^4} \label{eq:m}.
\end{equation}
\end{theorem}
\begin{IEEEproof}
See Appendix~\ref{app:2}.
\end{IEEEproof}
Thus, the transmission through the LRS is statistically equivalent to a direct transmission over a Nakagami fading channel.\footnote{The magnitude of the fading coefficient has a Nakagami distribution, although its phase is not uniformly distributed as is sometimes assumed for Nakagami fading. However, this is irrelevant, since we consider phase coherent detection.} The equivalent channel has $n^2$ larger receive power \emph{relative to the transmission through one reflector}. The average single-reflector SNR $\gamma_0$ can be very low because of the attenuation of the constituent channels and reflection. Next, we make a few remarks. 
\begin{remark}
The full distributions of the $2n$ coefficients $H_{i1}$, $H_{i2}$, $i=1,\ldots,n$, are irrelevant to the distribution of $|H|$, as this depends only on the average magnitudes~\eqref{eq:absH}. The individual coefficients may even have different distributions, as long as they are mutually independent and have the respective means.
\end{remark}
Furthermore, the phase errors manifest in~\eqref{eq:pdf_absH} through the first two trigonometric moments~\eqref{eq:trig_mom} of their common distribution, which are subunitary reals for the considered zero-mean symmetric distributions. 
\begin{remark}
The average SNR,  $\bar{\gamma}=n^2\gamma_0\E[|H|^2]=n^2\gamma_0\varphi_1^2 a^4<n^2\gamma_0$, is attenuated by $\varphi_1^2$, such that the broader the phase distribution (larger errors), the lower $\bar{\gamma}$.
\end{remark}
The average SNR does not grow indefinitely with $n$, because it cannot be larger than the transmit signal power divided by the receiver noise.\footnote{The upper bound would correspond to the contrived case when the emitted waves all arrive at the receiver without any absorption.}
\begin{remark}\label{remark3}
The quantity $n^2\gamma_0$ is $O(1)$ w.r.t. $n$. 
\end{remark}

\section{Performance Analysis}
We investigate how the performance of the system is impacted by the system size and phase errors. We consider the following types of errors:
\begin{itemize}
    \item \emph{phase estimation errors}: $\Theta_i$ is modelled as a zero-mean von Mises variable whose concentration parameter $\kappa$ captures the accuracy of the estimation. In this case, the characteristic function is $\varphi_p = \frac{I_p(\kappa)}{I_0(\kappa)}$, where $I_p$ is the modified Bessel function of the first kind and order $p$~\cite{MardiaJupp2000}.
    \item \emph{quantization errors}: when only a discrete set of $2^q$ phases can be configured, $q\geq 1$, the error $\Theta_i$ is assumed to be uniformly distributed over $[-2^{-q}\pi,2^{-q}\pi]$. For this distribution, we obtain $\varphi_1=\frac{\sin(2^{-q}\pi)}{2^{-q}\pi}$ and $\varphi_2=\frac{\sin(2^{-q+1}\pi)}{2^{-q+1}\pi}$. 
\end{itemize}
We study these types of errors separately, although the case with both estimation and quantization errors can be easily treated because, assuming they are independent, the resulting $\varphi_1$ and $\varphi_2$ are given by the product of the respective trigonometric moments. 

\subsection{Distribution of the SNR}
The performance of the system is fundamentally determined by the distribution of the instantaneous SNR, which equals $n^2\gamma_0 |H|^2$. 
\begin{corollary}
For large $n$, the instantaneous SNR, $n^2\gamma_0 |H|^2$, is gamma distributed. Its pdf can be expressed as
\begin{equation}\label{eq:pdf_SNR}
    f_{\mathrm{snr}}(\gamma) = \frac{m^m}{\Gamma(m)\bar{\gamma}^m} \gamma^{m-1}\exp\left(-\frac{m}{\bar{\gamma}}\gamma\right),
\end{equation}
where $m$ is given by~\eqref{eq:m} and $\bar\gamma = n^2 \varphi_1^2 a^4 \gamma_0$ is the average SNR.
\end{corollary}
\begin{IEEEproof}
The results follows from the fact that $|H|^2$ has a gamma distribution for $n$ large, see~\eqref{eq:cgf_approx2}, and the scaling factor $n^2\gamma_0$ is $O(1)$, as pointed out in Remark~\ref{remark3}. 
\end{IEEEproof}

\begin{figure}
    \centering
    \includegraphics[width=\columnwidth]{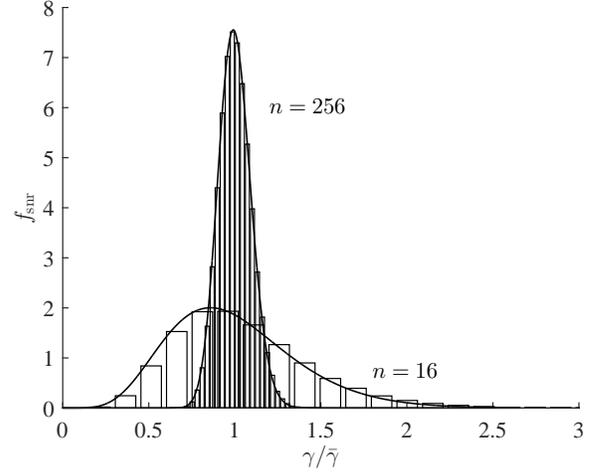}
    \caption{The pdf of the instantaneous SNR for $n=16$ and $n=256$, Rician fading on the source-LRS channels, Rayleigh fading on the LRS-destination channels and von Mises distributed phase errors. The normalized histograms are obtained through Monte Carlo simulation with $10^5$ trials, whereas the solid curves represent the gamma pdf~\eqref{eq:pdf_SNR} .}
    \label{fig:pdf_H2}
\end{figure}

As an example, the pdf~\eqref{eq:pdf_SNR} is displayed in Fig.~\ref{fig:pdf_H2} for the following scenario: the phase errors have a zero-mean von Mises distribution with concentration $\kappa=8$; the $\mathrm{S}$-to-$\mathrm{R}_i$ channels exhibit Rician fading with unit power and a Rice factor of $K=1$, which gives $a_1=\sqrt{\frac{\pi}{4(K+1)}}\,{}_1F_1(-1/2;1;K)$, while the fading over the $\mathrm{R}_i$-to-$\mathrm{D}$ channels is Rayleigh with unit power, for which $a_2=\sqrt{\pi}/2$. We observe a very good agreement between the theoretical approximation and the Monte Carlo estimate even for a moderate value of $n$. Also, when $n$ becomes large, the pdf concentrates around the mean $\bar{\gamma}$ and approaches a Gaussian pdf. 

\subsection{Average error probability}
\begin{figure*}
    \centering
    \subfloat[]{
    \includegraphics[width=\columnwidth]{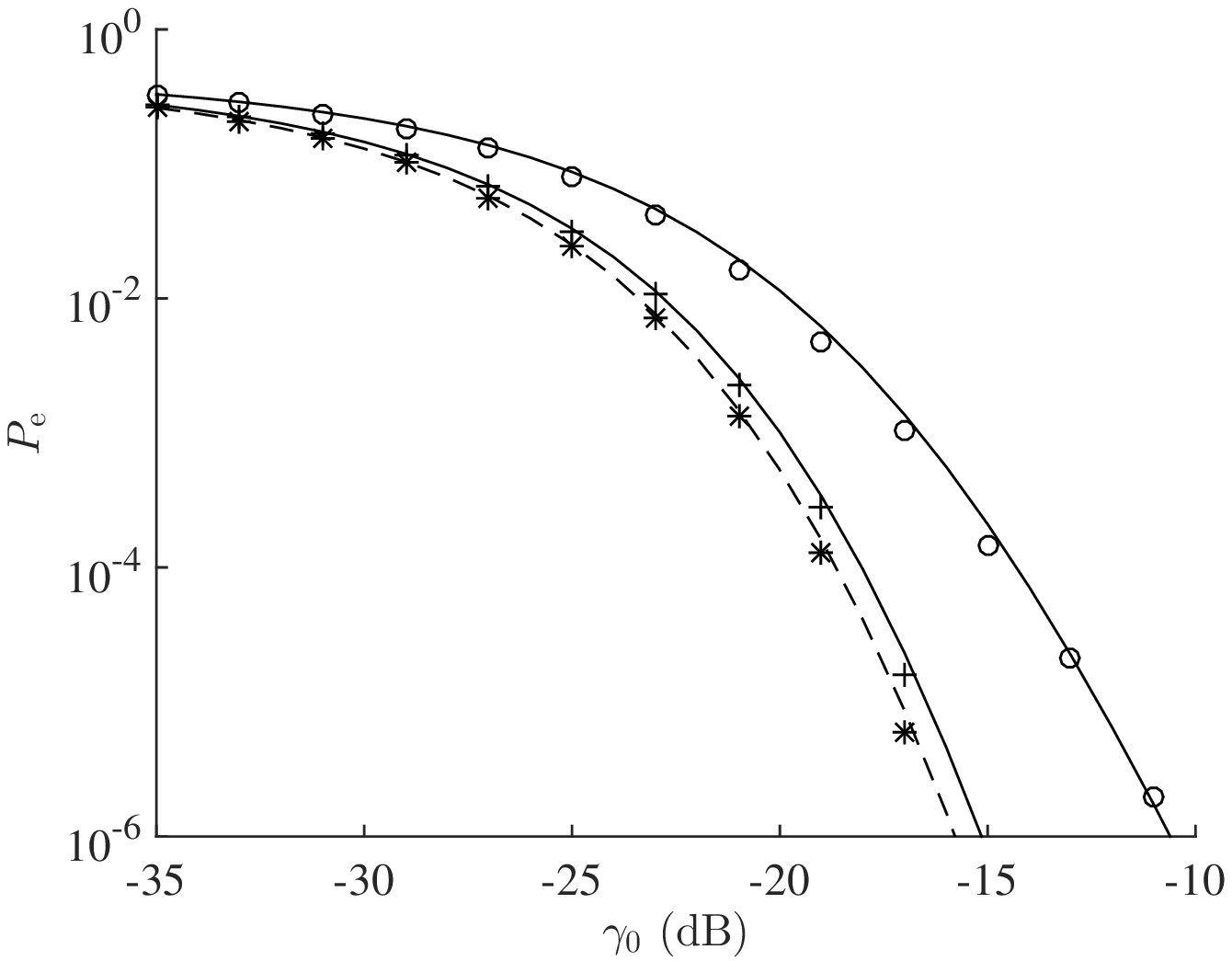}
    \label{fig:Pe_est}
    }
    \subfloat[]{
    \includegraphics[width=\columnwidth]{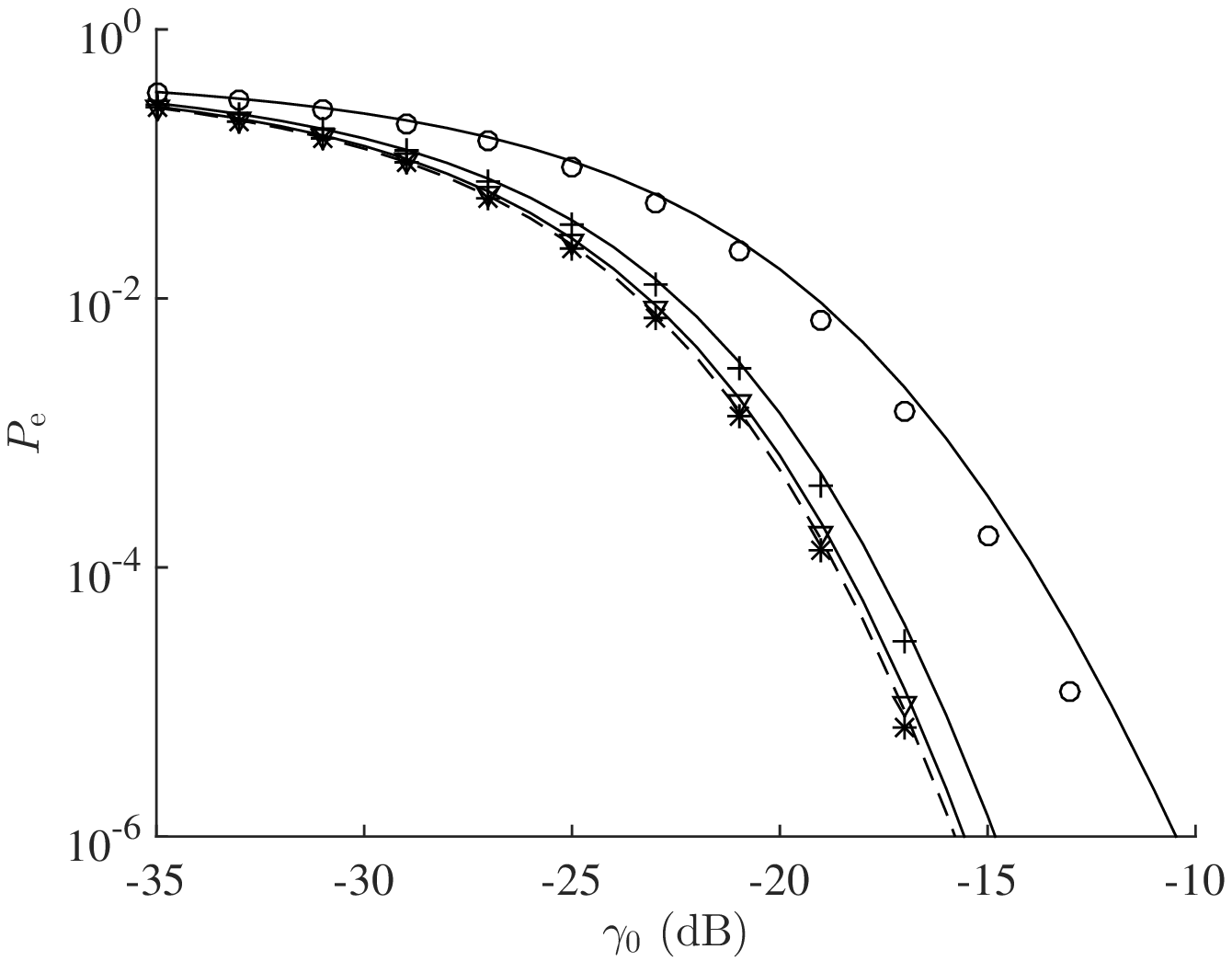}
    \label{fig:Pe_quant}
    }
    \caption{Average probability of error for $n=32$ under phase estimation errors (left) and quantization errors (right). The curves correspond to the theoretical performance for the Nakagami channel in Theorem~\ref{th:absH}, whereas the marked points are obtained via Monte Carlo simulations of the LRS system with $10^7$ trials. The dashed curves represent the ideal performance with zero errors. Left: the concentration parameter of the von Mises distribution is $\kappa=2$ and $\kappa=8$, in increasing order of performance. Right: the number of quantization bits is $q=1$, $2$ and $3$ bits, in increasing order of performance.}
    \label{fig:Pe}
\end{figure*}

We evaluate through simulations the average error probability, $P_\mathrm{e}$, of the LRS system for BPSK transmission, $n=32$ and the same fading models as in the previous example. In Fig.~\ref{fig:Pe}, the results for estimation and quantization errors are plotted, together with the theoretical performance~\cite{Simon2000} of the Nakagami fading channel given by Theorem~\ref{th:absH}. We observe a good agreement between the simulations and the theoretical prediction even for a limited $n$. Other results not included here show that the match is closer for $n=64$, whereas for $n=16$ the theoretical approximation is less accurate, although fairly good. As illustrated in Fig.~\ref{fig:Pe_est}, even a vague knowledge of the ideal phase shifts can be beneficial: e.g., when $\kappa=2$ (for which the distribution is broad) the gap to ideal performance is of about $5$ dB, while for $\kappa=8$ (for which the distribution is still not very concentrated) $P_\mathrm{e}$ is within $1$ dB from the ideal case. It is also remarkable to observe in Fig.~\ref{fig:Pe_quant} that phase quantization of just one bit is already at about $5$ dB from the ideal $P_\mathrm{e}$, while with two bits the error probability becomes close to the ideal. 

The average error probability for the Nakagami channel at high average SNR $\bar{\gamma}$ is~\cite{Wang2002}
\begin{equation}
    P_\mathrm{e}\approx \frac{m^{m-1}\Gamma(m+\frac{1}{2})}{2\sqrt{\pi}\Gamma(m)} \bar{\gamma}^{-m}.
\end{equation}
Using $\bar{\gamma} = n^2 \varphi_1^2 a^4 \gamma_0$, we express $P_\mathrm{e}(\gamma_0)=(G_\mathrm{c} \gamma_0)^{-G_\mathrm{d}}$, such that the diversity and coding gains w.r.t. the average single-reflector SNR $\gamma_0$ are $G_\mathrm{d} = m$ and $G_\mathrm{c} = n^2 \varphi_1^2 a^4\left[ \frac{m^{m-1}\Gamma(m+\frac{1}{2})}{2\sqrt{\pi}\Gamma(m)}\right]^{-1/m}$. These relationships may be useful to find the number of reflector elements required to achieve a certain $G_\mathrm{d}$ or $G_\mathrm{c}$, given the distribution of the phase errors (e.g., $\kappa$ for estimation errors). While $n$ can be easily established in the case of the diversity gain based on~\eqref{eq:m}, the relationship is more complicated in the case of the coding gain. 

\section{Conclusion}
We showed that the composite channel through an LRS with phase errors is equivalent to a point-to-point channel with Nakagami fading whose parameters are affected by the phase uncertainty through the first two trigonometric moments of the phase error distribution. Despite the phase shift errors, the average SNR still grows with $n^2$ (w.r.t. the single-reflector SNR) and the diversity order grows linearly with $n$; however, both are attenuated by the phase uncertainty. Also, the distributions of the fading coefficients of the two constituent channels manifest in the equivalent channel only though their average magnitudes. Numerical results for a limited number of reflectors showed good agreement with the theoretical prediction and that the performance is remarkably robust against the phase errors.

%Even though the direct S-D link was not included in our study, the analysis extends straightforwardly. In that case, the ideal reflector phases should align the phases of the reflected signals with the phase over the direct link. Thus, the equivalent fading coefficient is the sum of the Nakagami variable derived in this paper and the magnitude of the coefficient of the direct link, weighted by the corresponding SNRs. Existing results for the sum of gamma variables (e.g., see~\cite{Moschopoulos1985,Alouini2001}) apply immediately. 

\appendices

\section{Proof of Lemma~\ref{lm:H}}
\label{app:1}

The $n$ complex variables $|H_{i1}||H_{i2}| e^{j\Theta_i}$, $i=1,\ldots,n$, are i.i.d. with common mean $\mu = a^2\varphi_1$, variance $\nu=1-a^4|\varphi_1|^2$ and pseudo-variance $\rho=\varphi_2 - a^4\varphi_1^2$ (see~\cite{Picinbono1996} for the second-order characterization of complex variables). According to the assumptions on the distribution of $\Theta_i$, the trigonometric moments $\varphi_1,\varphi_2 \in\mathbb{R}$ and, consequently, $\mu,\rho\in\mathbb{R}$. By invoking the central limit theorem, we approximate the distribution of $H$ in~\eqref{eq:H} for $n$ large by a complex normal distribution, $\mathcal{CN}(\mu,\nu/n,\rho/n)$. Let $U=\Re(H)$ and $V=\Im(H)$. We have~\cite{Picinbono1996}
\begin{equation}
    \Cov[U,V] = \frac{1}{2}\Im\left(-\frac{\nu}{n}+\frac{\rho}{n}\right) = 0.
\end{equation}
Being jointly Gaussian with zero covariance, it follows that $U$ and $V$ are independent. Furthermore, $U\sim\mathcal{N}(\mu,\sigma_U^2)$ and $V\sim\mathcal{N}(0,\sigma_V^2)$ with $\mu = a^2\varphi_1$,
\begin{align}
    \sigma_U^2 &=\frac{1}{2}\Re\left(\frac{\nu}{n}+\frac{\rho}{n}\right) = \frac{1}{2n}(1+\varphi_2-2a^4\varphi_1^2)\\
    \sigma_V^2 &=\frac{1}{2}\Re\left(\frac{\nu}{n}-\frac{\rho}{n}\right) = \frac{1}{2n}(1-\varphi_2).
\end{align}

\section{Proof of Theorem~\ref{th:absH}}
\label{app:2}

We first consider the squared magnitude $|H|^2 = U^2 + V^2$. The normalized variable $U^2/\sigma_U^2$ has a non-central chi-squared distribution, while $V^2$ is gamma distributed with shape $1/2$ and scale $2\sigma_V^2$. Thus, $|H|^2$ is the sum of a \emph{scaled} non-central chi-squared variable and a gamma variable. We characterize the distribution of $|H|^2$ through its cumulant generating function, $K_{|H|^2}(t) = \ln \E[e^{t|H|^2}]$. Since $U$ and $V$ are independent (Lemma~\ref{lm:H}),
\begin{multline}\label{eq:cgf}
    K_{|H|^2}(t) = K_{U^2}(t) + K_{V^2}(t) = \frac{\mu^2 t}{1-2\sigma_U^2 t}\\ - \frac{1}{2}\ln(1-2\sigma_U^2 t) - \frac{1}{2}\ln(1-2\sigma_V^2 t)
\end{multline}
We now approximate for large $n$ the first term in the r.h.s. of~\eqref{eq:cgf}. We develop its Maclaurin series expansion as
\begin{multline*}
    \frac{\mu^2 t}{1-2\sigma_U^2 t} = \mu^2 t + \mu^2 (2\sigma_U^2) t^2 + \mu^2 (2\sigma_U^2)^2 t^3 +\ldots\\
    = \frac{\mu^2}{4\sigma_U^2}\sum_{k=1}^\infty \frac{k}{2^{k-1}} \, \frac{(4\sigma_U^2 t)^{k}}{k}
    = \frac{\mu^2}{4\sigma_U^2} \underbrace{ \sum_{k=1}^\infty \, \frac{(4\sigma_U^2 t)^{k}}{k} }_{=-\ln(1-4\sigma_U^2 t)} - g(t) 
\end{multline*}
where
\begin{align*}
    g(t) &= \frac{\mu^2}{4\sigma_U^2}\sum_{k=3}^\infty \left(1-\frac{k}{2^{k-1}}\right) \frac{(4\sigma_U^2 t)^{k}}{k}\\
    &< \frac{\mu^2}{4\sigma_U^2}\sum_{k=3}^\infty \frac{(4\sigma_U^2 t)^{k}}{k} = 4 \mu^2 \sigma_U^2 t^2\sum_{k=1}^\infty \frac{(4\sigma_U^2 t)^{k}}{k+2}\\
    &< 4 \mu^2 \sigma_U^2 t^2\sum_{k=1}^\infty \frac{(4\sigma_U^2 t)^{k}}{k} = -4 \mu^2 \sigma_U^2 t^2 \ln\left(1-4\sigma_U^2 t\right)
\end{align*}
Given that $\sigma_U^2=O(n^{-1})$, we have $g(t)=O(n^{-2})$, such that~\eqref{eq:cgf} is well approximated by\footnote{Alternatively, the Maclaurin expansion of~\eqref{eq:cgf} can be truncated to second-order, which corresponds to a Gaussian distribution and incurs an error of $O(n^{-2})$ as well. However, for moderate values of $n$, the Gaussian approximation is less accurate than the one we developed.}
\begin{multline}\label{eq:cgf_approx}
    K_{|H|^2}(t) = -\frac{\mu^2}{4\sigma_U^2}\ln(1-4\sigma_U^2 t) \\  -\frac{1}{2}\ln(1-2\sigma_U^2 t) - \frac{1}{2}\ln(1-2\sigma_V^2 t)
\end{multline}

Expression~\eqref{eq:cgf_approx} corresponds to the cumulant generating function of the sum of three independent gamma variables with (shape, scale) parameters $(\frac{\mu^2}{4\sigma_U^2},4\sigma_U^2)$, $(1/2,2\sigma_U^2)$ and respectively $(1/2,2\sigma_V^2)$. While the resulting distribution can be characterized based on~\cite{Moschopoulos1985}, we simplify by making a further approximation,
\begin{equation}\label{eq:cgf_approx2}
    K_{|H|^2}(t) = -\frac{\mu^2}{4\sigma_U^2}\ln(1-4\sigma_U^2 t),
\end{equation}
which has $O(n^{-1})$ error. 
Thus, for large $n$, $|H|^2$ has a gamma distribution with shape $\frac{\mu^2}{4\sigma_U^2}$ and scale $4\sigma_U^2$. It follows through variable transformation that $|H|$ has a Nakagami distribution with fading parameter (shape) 
\begin{equation*}
    m = \frac{\mu^2}{4\sigma_U^2} = \frac{n}{2}\,\frac{\varphi_1^2 a^4}{1 + \varphi_2 - 2\varphi_1^2 a^4}
\end{equation*}
and spread $\mu^2$.

\bibliographystyle{IEEEtran}
\bibliography{references.bib}

\end{document}